# Modifying the U-Net's Encoder-Decoder Architecture for Segmentation of Tumors in Breast Ultrasound Images

S. Derakhshandeh [a], A. Mahloojifar [a,*]

[a] *Department of Computer and Electrical Engineering, Tarbiat Modares University, Tehran, Iran*

**Abstract**

Segmentation is one of the most significant steps in image processing. Segmenting an image is a technique that makes it possible to separate a digital image into various areas based on the different characteristics of pixels in the image. In particular, segmentation of breast ultrasound images is widely used for cancer identification. As a result of image segmentation, it is possible to make early diagnoses of a diseases via medical images in a very effective way. Due to various ultrasound artifacts and noises, including speckle noise, low signal-to-noise ratio, and intensity heterogeneity, the process of accurately segmenting medical images, such as ultrasound images, is still a challenging task. In this paper, we present a new method to improve the accuracy and effectiveness of breast ultrasound image segmentation. More precisely, we propose a Neural Network (NN) based on U-Net and an encoder-decoder architecture. By taking U-Net as the basis, both encoder and decoder parts are developed by combining U-Net with other Deep Neural Networks (Res-Net and MultiResUNet) and introducing a new approach and block (Co-Block), which preserve as much as possible the low-level and the high-level features. Designed network is evaluated using the Breast Ultrasound Images (BUSI) Dataset. It consists of 780 images and the images are categorized into three classes, which are normal, benign, and malignant. According to our extensive evaluations on a public breast ultrasound dataset, designed network segments the breast lesions more accurately than other state-of-the-art deep learning methods. With only 8.88M parameters, our network (CResU-Net) obtained 82.88%, 77.5%, 90.3%, and 98.4% in terms of Dice similarity coefficients (DSC), Intersection over Union (IoU), Area under curve (AUC), and global accuracy (ACC), respectively, on BUSI dataset.

*Keywords*: Breast ultrasound image, Tumor segmentation, Deep neural networks, Concatenate block (Co-Block)

## 1. Introduction

There is no doubt that breast cancer is a deadly disease that affects humans all over the world. As a matter of fact, breast cancer is the leading cause of death for women in society at the moment [1]. In addition to improving the quality of life, early detection is also crucial to ensuring the survival of the individual. In the case of breast cancer, breast ultrasound imaging (BUS) is a reliable, low-cost, non-invasive, efficient, and popular modality to detect breast cancer at an early stage [2]. However, ultrasound images have a low quality and the resulting image highly suffers from speckle noise. Consequently, computer-aided diagnosis and accurate detection of breast cancer is a challenging task. The image segmentation is one of the most significant steps in image processing [3]. In addition, it helps to distinguish between areas with and without lesions, as well as irregularly shaped lesions in the breast [4]. There are various types of noise that appear in ultrasound images, particularly in breast ultrasound images, as discussed earlier. It is, therefore, difficult to manually extract information from these images and segment ultrasound images. In addition, radiologists and doctors are required to have a high level of experience, skill, and ability [5]. Thus, in order to automatically segment the images, it is necessary to use different models that are capable of automatically segmenting the images. Alternatively, automated segmentation is more convenient for doctors because it saves their time. To aid doctors in making correct and early diagnoses, these methods and models must be highly precise and accurate.

It has recently been shown that deep learning models, such as Convolutional Neural networks (CNNs) [6] can achieve exceptional results in the fields of computer vision, computer-aided diagnosis, and image processing, including segmentation, detection of objects, and classification. U-Net [7] has proven to be one of the most reliable CNN models for image segmentation compared to other CNN models. Different variants of U-Net have been used in the field of image segmentation. For example, DC-UNet [8] is one of the influential U-Net models in which an architecture is designed to replace the encoder-decoder. Quan Zhou et al. [9] proposed LAED-Net, which reduces network parameters significantly. They designed a lightweight decoder block. Another effective U-Net model is NAT-Net [10] which is an end-to-end noisy annotation tolerance network. Ilesanmi et al. [2] proposed an end-to-end network inspired by the U-Net architecture, which has a multi-pooling and double-concatenated convolution technique. A significant amount of deep learning models, as well as U-Net that have been designed to segment medical images, are encoder-decoder models. The encoder part uses convolutional and down-sampling layers. During the convolution process, various features are extracted from the input image. Also, during the down-sampling process, these features are selected and filtered as well. A set of selected features is then produced as outputs, which are referred to as dense tensors. The decoder part consists of convolution and up-sampling layers. During the decoding process, the dense tensor is used as input. To this end, it passes through layers of convolution and down-sampling. The predicted tensor is calculated through these layers, and the image is accordingly segmented.

[*]Corresponding Author: Ali Mahloojifar, Department of Biomedical Engineering, Faculty of Electrical and Computer Engineering, Tarbiat Modares University, Jalal Ale Ahmad, P.O. Box 14115-111, Tehran, Iran; Email: mahlooji@modares.ac.ir; Phone:+98(21)40338828.

As far as accuracy is concerned, encoder-decoder models have been found to be much better than traditional methods such as watershed-based models, clustering-based models, and threshold-based models, however there are still certain issues that need to be addressed, such as:

1) As a result of using down-sampling, low-level features that contain a lot of information will be lost. This is while such information will allow us to achieve a better level of segmentation accuracy as a result.
2) In the case of down-sampling, local information related to pixels in the image will be lost.

The above-mentioned challenges have been investigated by researchers and different methods have been developed so far. In this paper, in order to further improve the performance of the neural network in terms of image segmentation, we propose a new method and apply some modifications in both the encoder and decoder parts of the encoder-decoder model. A Neural Network (NN) is designed that has an encoder-decoder architecture and it is based on the U-Net architecture. Using U-Net as a foundation, both the encoder and decoder parts are modified by combining U-Net with other Deep Neural Networks (DNNs) and introducing a new approach that maintains the low-level features as much as possible. Our NN is evaluated on the breast ultrasound images (BUSI) dataset [11]. It contains 780 breast ultrasound images and is categorized into three classes: normal, benign, and malignant. Our main contributions are summarized below:

- An encoder-decoder architecture is proposed that combines different parts of deep NNs with the encoder-decoder U-Net architecture. In order to achieve a higher level of accuracy, we developed Co-block While U-Net and Res-Net have been widely applied in medical image segmentation, our proposed method introduces a novel *Co-Block* that uniquely combines low- and high-level feature retention. The Co-block extracts more complex features from the input images by increasing the number of convolutional layer filters, also maximizes the use and preservation of the extracted features by employing the concatenate operator. As part of the network, both the encoder part as well as the decoder part utilizes these blocks. Using the newly added blocks allow us to retain useful features and pass them along to the next block as inputs.
- In order to save low-level features, a new neural network architecture is proposed. A reliable level of accuracy can be gained by modifying both the encoder and decoder parts by using the Co-block and Res-Net's identity/MultiResUNet's blocks to achieve the desired results.
- In the proposed method, we tried to increase accuracy by using fewer features. Due to this, the complexity of the computation has been reduced.
- Proposed NN is tested on the BUSI dataset. Various metrics have been used to evaluate designed network, including dice score (DSC), global accuracy (ACC), intersection-over-union (IoU), and area under curve (AUC). The experiments demonstrate that Proposed NN achieves reliable accuracy and effectiveness.

The paper is organized as follows. Section 2 overviews the related works. Section 3 describes the method and details of designed neural network architecture and dataset. Experimental results are provided in Section 4. Finally, the results and conclusion presented in Section 5 and Section 6, respectively.

## 2. Related works

In computer vision, different techniques are available to perform image segmentation which can be categorized into two main parts: deep learning algorithms and traditional algorithms such as thresholding, edge detection, and watershed-based algorithms [3].
Researchers are using two kinds of algorithms to segment images in different ways. A brief explanation about each category is presented in the following.

## 2.1 Classical algorithms

*2.1.1 Thresholding-based method*

Thresholding is one of the widely used techniques for monochromatic image segmentation and is used for BUSI segmentation [12]. As the simplest image segmentation method, thresholding is based on a clip level (or threshold value) to convert a grayscale image into a binary image. The thresholding-based method is relatively simple. However, it may not work well for breast ultrasound images, since this method only considers gray-level statistics, and does not take spatial information into consideration. In addition, in some images, there is a certain overlap ratio in the gray level distribution between the object and the background. From 2001 to 2004, Horsch et al. [13] focused on threshold segmentation of BUSI and proposed improved algorithms.

*2.1.2 Watershed-based & Clustering-based methods:*

In Watershed-based method, the gradient magnitude of an image is considered as a topographic level. This is one of the most popular image segmentation algorithms for grayscale images [14], [15]. It is believed that the pixels with the highest gradient intensity range correspond to the watershed lines that show the boundaries of the region. Due to the sensitive nature of this algorithm, it can easily over-segment images when there is noise present. Clustering-based is a method to perform Image Segmentation of pixel-wise segmentation. In this type of segmentation, an attempt is made to cluster the pixels that are together [4].

## 2.2 Neural networks

(NN)-based segmentation methods, which transform the segmentation problem into a classification decision based on sets of input features, are popular and have been proven to be very accurate [16], [17]. In 1999, Binder et al. [18] investigated echocardiographic image segmentation using an artificial NN. A deep learning method is a new research direction in machine learning and artificial intelligence. Using deep neural networks, it simulates how the brain learns as well as extracts features from large-scale data sets without requiring any supervision. As part of the diverse family of deep learning models, Convolutional Neural Networks (CNN) have demonstrated excellent performance over a range of tasks in computer vision such as image classification, object recognition, and many others. CNNs are a class of NN produced by integrating deep learning technologies with image processing technologies, which is an interesting application. It started with the discovery of Hubel and Wiesel [19]. They explained that there are simple and complex neurons in the primary visual cortex and visual processing always starts with simple structures such as directional edges. Among different CNN models, U-Net has shown significant performance. It has proved to be an excellent computer vision tool because of its high performance, and as a result it has been widely used in various aspects of computer vision. For example, the attention-enriched deep learning model [20] proposes integrating visual saliency into a deep learning model for BUSI segmentation. Another effective model of U-Net is DoubleU-Net [21]. It is a combination of two U-Net models. The encoder part of the first one is a pre-trained VGG-19 [22] model, and then they added another U-Net at the end. Using this model, more semantic information can be obtained. Zhou z et al. [23] proposed a U-Net++ model; its architecture is based on U-Net and has a dense connection between the encoder and decoder in every stage. Trans-UNet [24] is a powerful model of U-Net that uses both Transformers and U-Net. This technique allows Trans-UNet to use better spatial details. There are many CNN models among which, the Res-Net model [25] is one of the most influential. It has been shown that residual networks are easier to be optimized compared to deep CNNs. As a result, using this framework, one can solve the problem of not being able to train deep CNNs. In order to achieve end-to-end pixel-level image segmentation, Seg-Net [26] builds an encoder-decoder symmetric structure based on FCN's semantic segmentation task. VGG-16 is used in the encoder to investigate object information. According to the object information for each pixel, the decoder produces the final image from the parsed details; each pixel has its own label. In terms of its architecture, the U-Net can be compared to Seg-Net's encoder-decoder structure. U-Net is an excellent tool for segmenting medical images for many reasons, mainly due to its structure which allows to combine low-level information with high-level information at the same time, which makes it ideal for segmenting medical images. As a whole, we aim to improve the accuracy of ultrasound medical image segmentation through the combination of U-Net encoder-decoder structure with other deep learning networks.

## 3. Method

*3.1 Dataset*

In order to conduct this study, the BUSI dataset was used. This dataset contains a mixture of normal breast ultrasound images and images with benign and malignant masses in the breast. An ultrasound system called the LOGIQ E9 and the LOGIQ E9 Agile were used at Baheya Hospital, Cairo, Egypt to acquire these images. Women between the ages of 25 and 75 were included in the study who had breast ultrasound images taken at baseline. It consists of 780 images with an average size of 500 × 500 pixels, which represents the size of the entire dataset. Furthermore, each image has its own ground truth image (mask image) that serves as the basis for each image. Table 1 shows the number of normal, benign and malignant images. Fig. 1 illustrates Samples of Ultrasound breast images and Ground Truth Images. With only 780 images, the BUSI dataset may not fully represent the diversity of real-world ultrasound imaging.

Table 1: The classes of breast cases and the number of images in each case.

| *Mass type* | *Number of Images* |
|---|---|
| *Benign* | 437 |
| *Malignant* | 210 |
| *Normal* | 133 |
| *Total* | 780 |

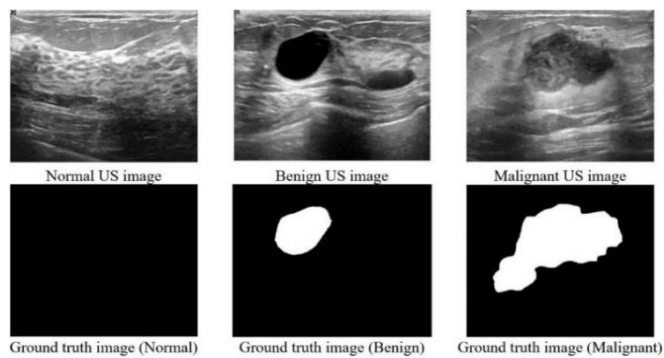

Figure 1: Samples of Ultrasound breast images and Ground Truth Images.

*3.2 Pre-processing*

According to earlier discussion, the image inputs have an average size of 500 × 500 pixels. The input image size is reduced to 256 × 256 pixels. As a result of reducing the size of the input image, computation becomes simpler, and consequently, the performance of the network will be improved. In addition, data augmentation is performed by reducing speckle artifacts and rotating the images of training data. The speckle artifacts in the images of training dataset has been reduced using a non-local means (NLM) method [27]. In order to reduce speckles artifacts in ultrasound images, NLM is widely used. It divides the ultrasound image into several blocks by overlapping them. By calculating the mean value of each block, it determines how similar the reference block (the search window) is to the remaining blocks of the image. Therefore, NLM can reduce artifacts not only near the selected block, but also far away. Also, the images of the training dataset were rotated 180 degrees for data augmentation.

*3.3 Methodology*

Fig. 2 illustrates the whole architecture of the proposed network. The general structure of the proposed network is based on the U-net encoder-decoder architecture in which both the encoder and decoder parts are modified. This is done by combining Res-Net and other blocks of deep networks into the encoder and decoder parts. As a part of its encoder, the U-Net algorithm consists of layers of convolutional and down-sampling algorithms that extract and filter features from the data. It is important to note that filtering the extracted features results in the loss of low-level features. One significant characteristic of the proposed network is its utilization of low-level features, bringing them up to the output layers before they are filtered. The information obtained from this method helps us to improve the accuracy of the segmentation process. Assume that $O_{ME}^i$ is the output of the ith Encoder block after max-poling layer, $O_E^i$ is the output of the ith Encoder block, and $O_D^i$ is the output of the ith Decoder block.

*3.3.1 Encoder*

The encoder part of proposed network is composed of six blocks (see Fig. 2). We have designed concatenate block (Co-Block) (see Fig. 3). In the following, it will be mentioned which encoder blocks are Co-blocks. During the Co-block, three convolutional layers are added, followed by 3 batch normalization layers, and 3 activation layers (ReLU function). In Co-block, the input fed into first convolution layer with kernel size of 3×3. Then, it passes through activation (ReLU function) and batch normalization layers. Again, the output features fed into a second convolution layer (with twice filters of the first convolutional layer), an activation layer, and a batch normalization layer. The output is then concatenated with the output of the first convolution layer. The concatenated features are then passed through a third convolution layer with quadruple filters, an activation layer (ReLU function), and a batch normalization layer. Finally, the output is concatenated with output of first concatenated features. In the proposed *Co-Block*, the concatenation operation not only retains features across layers but also allows effective integration of multi-scale information. Unlike conventional encoder blocks, which risk information loss through sequential operations, the Co-Block ensures minimal feature degradation by combining outputs at multiple stages The Co-block offers a significant advantage by increasing the depth of feature extraction in input images. This is achieved through the expansion of convolutional layer filters, resulting in the extraction of more complex features. Additionally, the Co-block maximizes the utilization and preservation of extracted features by employing the concatenate operator. The input image has a 256×256 size that is an input for the first block. The first, second, and fourth E-Blocks consist of designed Co-Blocks. Also, Res-Net's identity block and MultiResUNet's block [28] (see Fig. 4 & 5) make up the third and fifth E-Blocks, respectively. A max-pooling layer (pool size = 2×2) is added after each E-Block. The identity block is the standard block used in Res-Nets and corresponds to the case where the input activation has the same dimension as the output activation. The convolutional layers extract features from the input image. Also, the max-pooling layers serve to filter the extracted features, keep high-level features still accessible, and reduce the size of the image. The first, second, third, 4[th], and 5[th] E-Block has 16, 32, 64, 128, and 256 filters for convolutional layers as an input, respectively. The last block has 2 convolution layers with 512 filters, a kernel size of 3×3, and a dropout layer. Also, the 5[th] block has a dropout layer. As mentioned before, proposed Co-Block allows us to use both low-level and high-level features. Low-level information helps to improve accuracy. Also, High-level information helps to extract complex features.

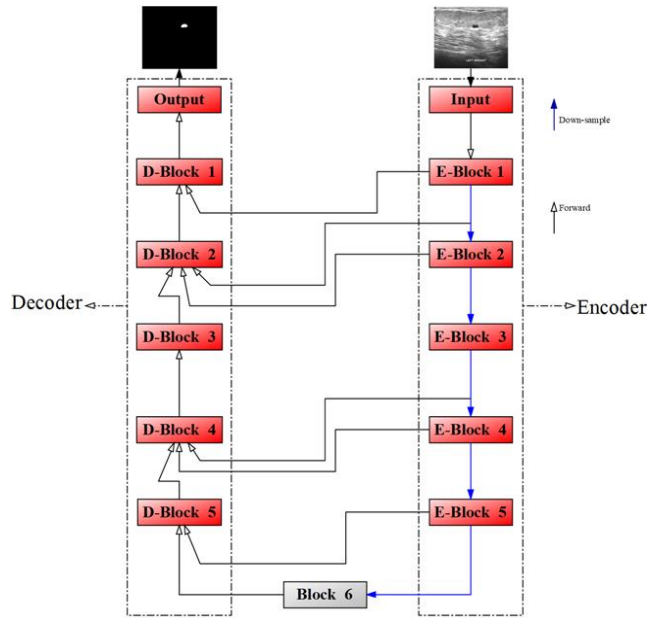

Figure 2: The overall network architecture of the proposed network (CResU-Net). The E-Blocks are a series of encoding modules, and the D-Blocks represent decoding modules. The black arrows denote information flow, and the blue arrows denote Down-sampling operation.

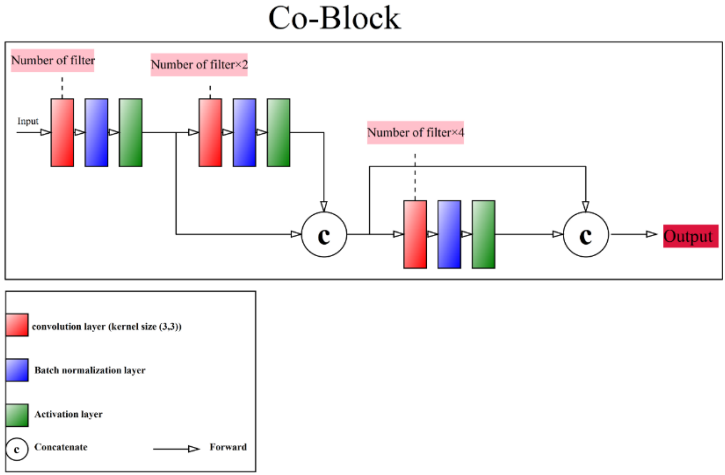

Figure 3: The description of the concatenate block. (The first, second, third, 4th, and 5th E-Block has 16, 32, 64, 128, and 256 filters for convolutional layers as an input, respectively.)

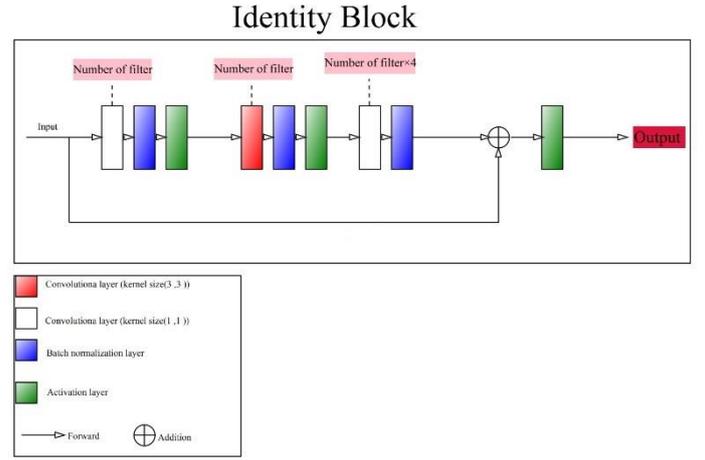

Figure 4: The description of the Indentity block. (The first, second, third, 4th, and 5th E-Block has 16, 32, 64, 128, and 256 number of filters for convolutional layers as an input, respectively.)

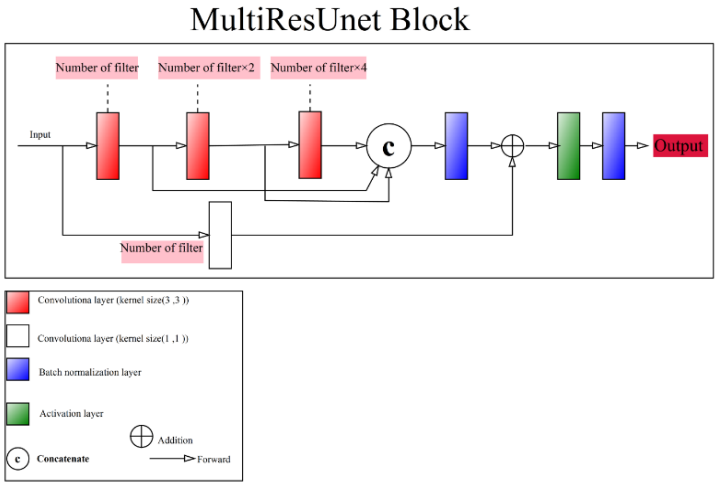

Figure 5: The description of the MultiResUnet block. (The first, second, third, 4th, and 5th E-Block has 16, 32, 64, 128, and 256 number of filters for convolutional layers as an input, respectively.)

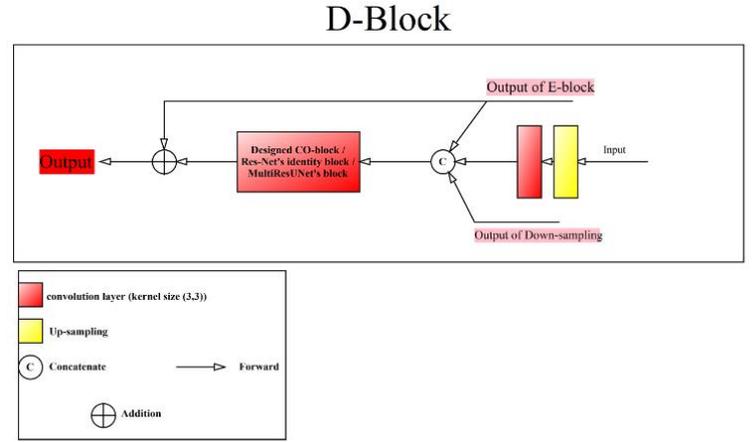

Figure 6: The description of the Decoder block (D-Block). (The first, second, third, 4th, and 5th E-Block has 16, 32, 64, 128, and 256 number of filters for convolutional layers as an input, respectively.).

*3.3.2 Decoder*

The whole architecture of the Decoder blocks is shown in Fig. 6. In order to make a more accurate segmentation prediction, the algorithm fuses and integrates high-level and low-level features. It is necessary for features to have the same dimensions in order to be integrated. It has already been mentioned that $O_E^i$ and $O_D^i$ are the outputs of the ith encoder and decoder blocks, respectively. During the Decoder block, an up-sampling layer is added, followed by a convolution layer. Then, the decoder block fuses the up-sampled $O_D^{i+1}$, $O_{ME}^{i-1}$, and $O_E^i$ by concatenation:

$$fb^i = concatenate(O_E^i, O_{ME}^{i-1}, O_D^{i+1}) \qquad (1)$$

Where $fb^i$ are features of ith block after concatenation. It is necessary to mention that D-Block 5 & 1 have two inputs ($O_E^i$ & $O_D^{i+1}$) and D-Block 3 has one input ($O_D^{i+1}$). Thereafter, the features are fed into the designed

Co-Block or Res-Net's identity/MultiResUNet's block, where each channel of the feature is separately convoluted. The final step is element-wise adding convolved $fb^i$ with $O_{CE}^i$:

$$O_D^i = fb^i \oplus O_{CE}^i \tag{2}$$

The next block will take $O_D^i$ as input. As compared to other blocks, there is a difference between the third and fifth D-blocks. Because of the matching dimensions of features in the third and fifth D-Blocks, we tend to replace the Co-Block with Res-Net's identity block/MultiResUNet's block instead of the Co-Block.

## 4. Results

*4.1 Metrics*

The BUSI dataset is used to evaluate the level of segmentation accuracy of designed network. Based on the results obtained from the research, it is clear that designed network achieves a reliable, as well as the highest level of segmentation accuracy. Dice score (DSC) and global accuracy (ACC) are used to evaluate performance of the model. The mentioned parameters are defined as below:

$$DSC = \frac{2|A_g \cap A_p|}{|A_g| + |A_p|} \tag{3}$$

$$ACC = \frac{|A_g \cap A_p| + |\overline{A_g} - A_g \cup A_p|}{|A_g| + |\overline{A_g}|} \tag{4}$$

Where $A_g$ refers to the number of pixels within a tumor region in the segmented images which are derived from ground truth images. In the ground truth segmented images, $\overline{A_g}$ represents pixels in the background area without tumors. Also, $A_p$ is defined as the set of pixels that, when segmented using a particular method, seems to correspond to a tumor region.

Other metrics that are used for performance evaluation are Intersection-over-Union (IoU), Precision, Recall, and Area under curve (AUC). In all of these metrics, higher values indicate better network performance. All of values are in the range of [0, 1]. The IoU evaluation metric is expressed according to the following equation:

$$\text{IoU} = J(A_g, A_p) = \frac{|A_g \cap A_p|}{|A_g \cup A_p|} \tag{5}$$

*4.2 Implementation*

The model is implemented using TensorFlow and Keras frameworks. The model is trained with a NVIDIA GeForce RTX 2080 GPU. The input images are all resized to $256 \times 256$ pixels for dataset. Images of the dataset were divided into training and testing images. Fivefold cross-validation is applied in the study, meaning the training set consisted of four-folds (80% of the images), and the testing set comprised one-fold (20% of the images). Validation during training is performed on 20% of the training set of images. The network is trained five times, and then the average mean and standard deviation of all metrics are calculated. There are 200 epochs in the training process of the model. The performance of the system is further improved by applying some data augmentation techniques. In order to train designed network, the Dice loss metric is used. Due to the imbalanced classes in our dataset, we prefer using the dice loss metric. Dice coefficient performs better at class imbalanced.

*4.3 Experimental results*

Fig. 7 shows the prediction results of U-Net, Seg-Net, D-Unet, and proposed model on the BUSI dataset. It can be seen that the shape of segmented tumors is better predicted by proposed model than other models. Also, by using the proposed method, the edges are detected more accurately compared to other models. Furthermore, in some images, there are two tumors located next to each other. Based on proposed model, we are able to correctly separate them from one another. The malignant tumor, on the other hand, often has irregular shapes and no clear edges or boundaries. Therefore, segmenting tumor is more difficult in such cases. A quantitative comparison between proposed and a variety of other models on BUSI dataset is given in Table 2. It can be concluded that proposed model has higher accuracy, lower computation and less parameters in comparison with other models.

The results indicate that proposed model outperforms U-Net in DSC, IoU, ACC, and AUC by 19.68%, 23.3%, 2.7%, and 5.8%, respectively. The segmentation accuracy of D-Unet is 63.6%, 57.8%, 96.0%, and 88.8% in terms of DSC, IoU, ACC, and AUC, respectively. As compared with D-Unet's model, proposed model results in a higher DSC, IoU, ACC, and AUC for about 19.28%, 19.8%, 2.4%, and 1.6%, respectively. As a result, proposed model is much more accurate at predicting and detecting edges than D-Unet.

There are three types of LAEDNet that are designed to segment ultrasound medical images. On the BUSI dataset, LAEDNet-L shows to be the most effective segmentation type among these types. There has been an improvement in the accuracy of LAEDNet-L's segmentation, however, in terms of DSC, IoU, and AUC, respectively, at 75.0 %, 67.6 %, and 91.3 %. Unfortunately, they did not provide any information about global accuracy. In terms of DSC, IoU, proposed model actually outperforms it by 7.88%, 9.9%, respectively.

There is no doubt that U-Net++ is one of the most effective models for segmenting medical images. The segmentation accuracy achieved by it can be defined as 70.9%, 63.8%, and 90.4% in terms of DSC, IoU, and AUC, respectively. The results are once again in favor of proposed model, especially when it comes to DSC, IoU, where it significantly outperforms its competitor by 11.98%, and 13.7%, respectively. The parameters of the networks are also measured in order to evaluate the size and computation of the networks. Compared to other networks, proposed model has a much smaller number of parameters than other models. Consequently, we are able to make proposed network lighter and with less computation, which results in a faster network. There is an optimal tradeoff between segmentation accuracy and computation time in the proposed method, making it the best option.

Table 2: Comparison results between state-of-the-art networks and proposed network on BUSI dataset.

| Model | DSC (%) | IoU (%) | AUC (%) | ACC (%) | Params (M) |
|---|---|---|---|---|---|
| U-Net | 63.2 ± 1.1 | 54.2 ± 3.9 | 84.5 ± 2.0 | 95.7 ± 0.3 | 28.97 |
| U-Net++ | 70.9 ± 1.8 | 63.8 ± 2.7 | 90.4 ± 1.3 | - | 36.15 |
| LAEDNet-L | 75.0 ± 0.9 | 67.6 ± 3.4 | **91.3 ± 1.1** | - | 63.25 |
| D-UNet | 63.6 ± 3.1 | 57.7 ± 3.0 | 88.7 ± 1.0 | 96.0 ± 1.9 | - |
| Seg-Net | 55.4 ± 3.1 | 47.5 ± 2.8 | 87.4 ± 3.0 | 95.5 ± 0.1 | 42.02 |
| **Proposed method** | **82.88 ± 3.1**↑ | **77.5 ± 2.9**↑ | 90.3 ± 1.3 | **98.4 ± 0.2**↑ | **8.88**↓ |

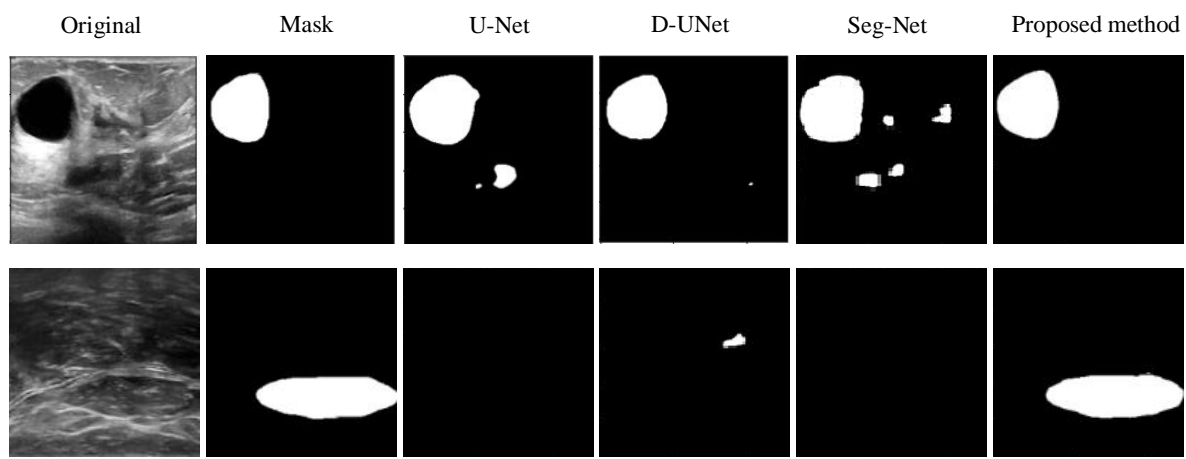

| Original | Mask | U-Net | D-UNet | Seg-Net | Proposed method |

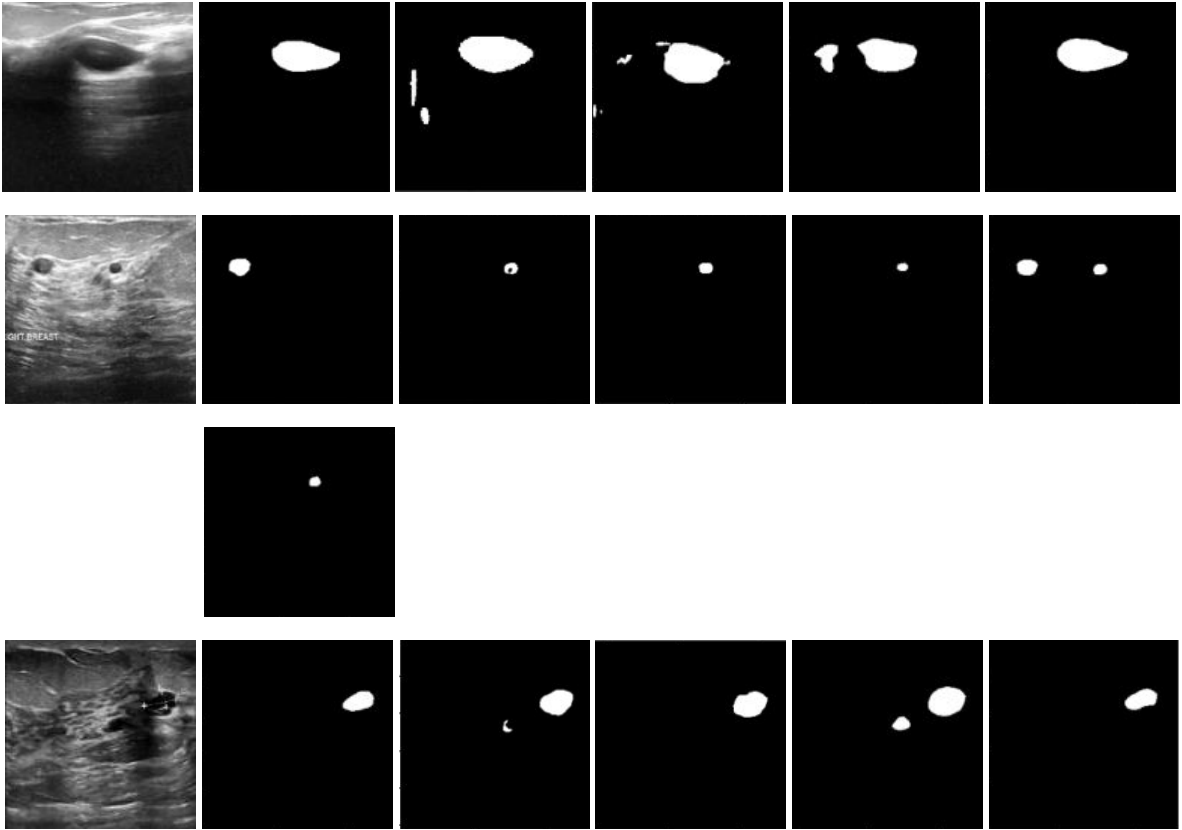

Figure 7: Some visual examples of qualitative segmentation outputs on BUSI dataset. From left to right are results of U-Net, D-UNet, Seg-Net, and proposed network. Note that the input image at forth row has two masses and two masks. Second mask is shown below the first mask.

Table 3: Ablation study results showing the impact of individual components (Res-Net Block, MultiResUNet Block, and Co-Block) on segmentation performance metrics (DSC, IoU, AUC, and ACC). The proposed model with all components achieves the highest perform, demonstrating the effectiveness of the Co-Block in integrating low- and high-level features.

| Configuration | DSC (%) | IoU (%) | AUC (%) | ACC (%) |
|---|---|---|---|---|
| U-Net Base | 63.2 | 54.2 | 84.5 | 95.7 |
| + Res-Net Block | 72.5 | 65.4 | 89.3 | 97.1 |
| + MultiResUNet Block | 78.4 | 70.6 | 90.1 | 97.8 |
| + Co-Block (Proposed Full Model) | 82.88 | 77.5 | 90.3 | 98.4 |

Table 4: Comparison of model parameters, memory usage, and inference time for various segmentation networks. The proposed model demonstrates significantly lower computational demands with only 8.88M parameters, 400 MB memory usage, and 45 ms/frame inference time, making it highly suitable for real-time applications compared to state-of-the-art models.

| Model | Parameters (M) | Memory Usage (MB) | Inference Time (ms/frame) |
|---|---|---|---|
| U-Net | 28.97 | 1200 | 100 |
| U-Net++ | 36.15 | 1400 | 120 |
| LAEDNet-L | 63.25 | 2400 | 160 |
| Seg-Net | 42.02 | 1800 | 140 |
| **Proposed** | **8.88** | **400** | **45** |

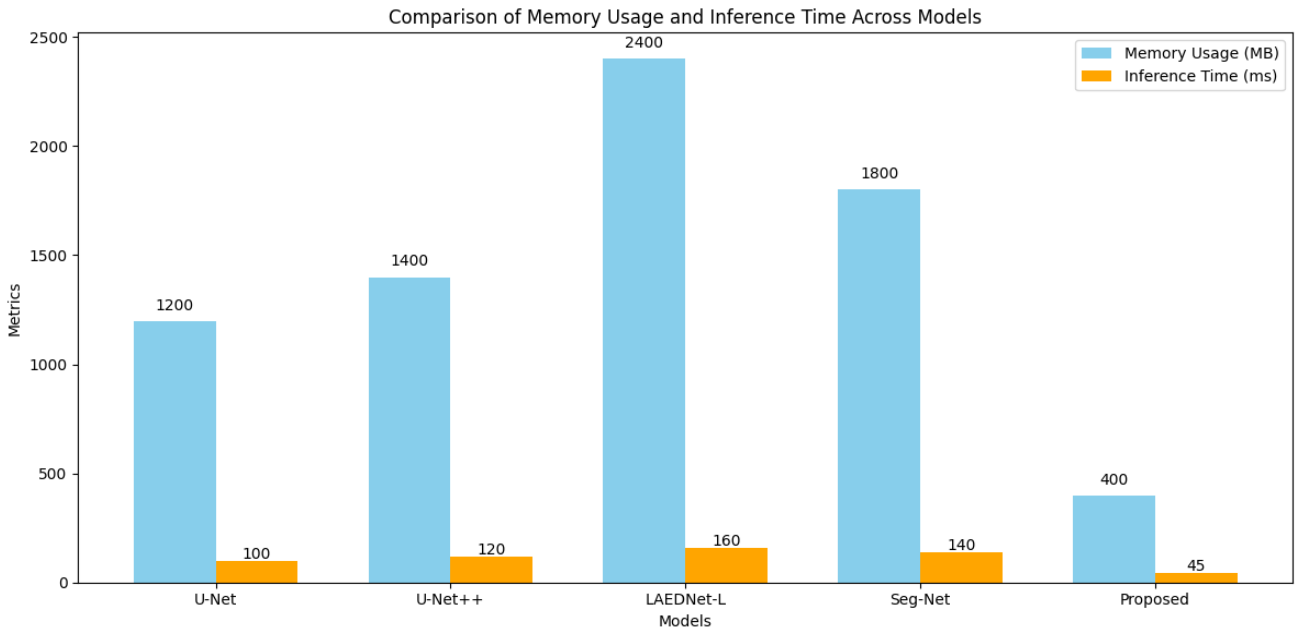

Figure 8: Bar chart comparing memory usage and inference time for different segmentation networks. The proposed model exhibits the lowest memory usage (400 MB) and fastest inference time (45 ms/frame), highlighting its computational efficiency relative to U-Net, U-Net++, LAEDNet-L, and Seg-Net.

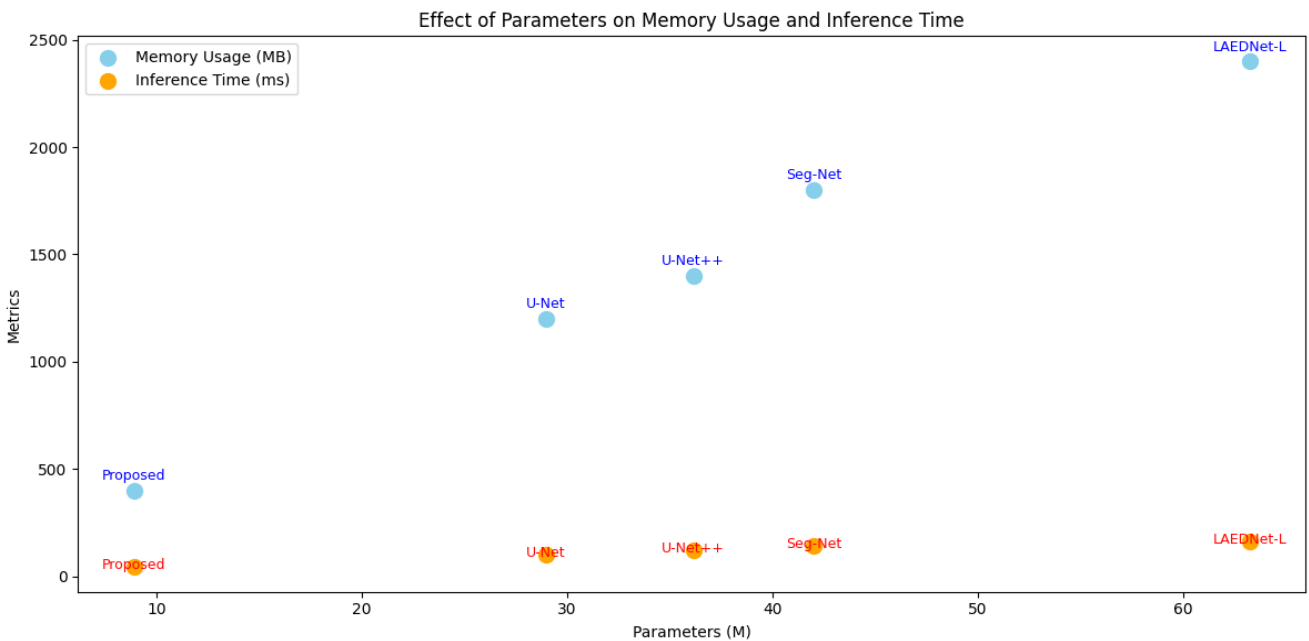

Figure 9: Scatter plot illustrating the relationship between model parameters, memory usage, and inference time. The proposed model, with the fewest parameters (8.88M), achieves the lowest computational demands, demonstrating a strong correlation between reduced parameter count and enhanced efficiency.

The **proposed model** uses significantly less memory (~400 MB) compared to U-Net (~1200 MB) and U-Net++ (~1400 MB). The inference time for the proposed model (~45 ms/frame) is also much faster than U-Net (~100 ms/frame) and U-Net++ (~120 ms/frame). The reduced memory usage indicates the proposed model is lightweight, making it suitable for deployment on devices with limited resources, such as portable ultrasound machines or embedded systems. LAEDNet-L has the highest memory usage (~2400 MB) and longest inference time (~160 ms/frame) among the models due to its large parameter count (63.25M). Seg-Net is better than LAEDNet-L in terms of memory usage (~1800 MB) and inference time (~140 ms/frame), but it still lags behind the proposed model. The **proposed model** exhibits a drastic reduction in both metrics: Memory usage is reduced by ~83% compared to Seg-Net and ~83.3% compared to LAEDNet-L. Inference time is ~3.5x faster than Seg-Net and ~3.6x faster than LAEDNet-L. LAEDNet-L and Seg-Net require significant computational resources, making them less suitable for real-time or resource-constrained applications.

The proposed model's lightweight design makes it highly efficient and ideal for real-time clinical use. Faster inference time makes it viable for real-time applications like live segmentation during medical procedures Models with more parameters (e.g., U-Net++ with 36.15M) show higher memory usage and longer inference times. The proposed model, with the fewest parameters (8.88M), demonstrates a substantial improvement in both metrics. The correlation between the number of parameters and memory/inference time indicates that parameter reduction directly impacts computational efficiency. This trade-off does not come at the cost of accuracy, as shown in the earlier ablation study and segmentation performance metrics. With ~45 ms/frame inference time, the proposed model can handle 22 frames per second, meeting real-time requirements for medical imaging.

The low memory usage (~400 MB) ensures compatibility with GPUs or CPUs that have limited VRAM, expanding the model's usability to a broader range of hardware. The compact model size (parameters: 8.88M) facilitates easier deployment in clinical environments, especially on edge devices or mobile platforms. While fewer parameters enhance efficiency, there is a potential trade-off in learning capacity. However, the proposed model mitigates this by carefully integrating the **Co-Block** and Res-Net components. The results are based on the BUSI dataset, so generalizing these findings to other datasets (e.g., those with larger images or more noise) requires additional testing.

In addition, using benign and malignant breast ultrasound images of BUSI, the robustness of the network in segmenting malignant and benign lesions is evaluated. The performance of the method has been compared with other state-of-the-art deep learning methods. They are U-Net, Att U-net [29], U-Net++, Seg-Net, RCA-IU-net [30], and RRCNet [31]. A quantitative comparison between proposed model and a variety of other models is given in Table 5. Four-fold cross-validation was performed on BUSI with benign and malignant images.

Table 5: Comparison results between state-of-the-art networks and Proposed network on Benign and Malignant lesions in BUSI dataset. Asterisks indicate that the difference between Proposed method and the competing method is significant using a paired student's t-test. (*:P<0.05).

| Methods | Malignant lesions of BUSI | | | | Benign lesions of BUSI | | | | Params (M) |
|---|---|---|---|---|---|---|---|---|---|
| | DSC (%) | IoU (%) | Precision (%) | Recall (%) | DSC (%) | IoU (%) | Precision (%) | Recall (%) | |
| U-Net (2015) | 63.47 ± 2.38 | 51.11 ± 2.62 | 64.96 ± 2.55 | 68.86 ± 4.27 | 70.49 ± 3.23 | 61.53 ± 3.98 | 74.97 ± 2.80 | 73.97 ± 5.81 | 28.97 |
| Att U-net (2018) | 62.95 ± 2.14 | 51.12 ± 2.35 | 61.62 ± 0.9 | 72.57 ± 2.17 | 73.30 ± 2.00 | 65.03 ± 2.05 | 75.24 ± 1.68 | 79.44 ± 2.84 | 35.56 |
| U-Net++ (2019) | 65.52 ± 2.75 | 54.03 ± 3.03 | 65.50 ± 2.94 | 73.43 ± 2.10 | 75.56 ± 2.79 | 68.25 ± 2.75* | 75.93 ± 3.66 | 81.58 ± 1.09* | 36.63 |
| Seg-Net (2017) | 65.90 ± 1.97 | 54.89 ± 1.78 | 63.79 ± 2.65 | 77.25 ± 4.02* | 75.47 ± 2.91 | 67.89 ± 3.31 | 76.96 ± 3.11 | 79.57 ± 2.21 | 42.02 |
| RCA-IU-net (2022) | 64.95 ± 3.06 | 53.07 ± 3.03 | 63.74 ± 3.34 | 74.96 ± 1.54 | 74.36 ± 2.30 | 66.50 ± 2.05 | 75.32 ± 3.09 | 80.93 ± 1.49 | 41.02 |
| RRCNet (2023) | 70.79 ± 2.29 | 59.95 ± 2.32 | 69.72 ± 1.60 | **77.63 ± 4.16** | 78.85 ± 2.29 | 71.83 ± 2.25 | 78.73 ± 3.38 | 83.37 ± 1.40 | 43.94 |
| **Proposed method** | **72.71 ± 1.31**↑ | **62.31 ± 2.26**↑ | **85.80 ± 1.65**↑ | 69.40 ± 3.11 | **92.42 ± 2.00**↑ | **86.13 ± 1.71**↑ | **91.80 ± 2.96**↑ | **93.74 ± 2.65**↑ | **8.88**↓ |

Proposed method achieves the best results for both benign and malignant BUSI segmentation, as shown in Table 5. Accordingly, proposed method has 94.4% DSC, 86.1% IoU, 91.8% precision, and 93.7% recall on benign BUSI. As compared to the second-best results, these metrics improved by 13.5%, 14.3%, 13.0%, and 10.3%, respectively. On the segmentation of malignant tumors, the results of proposed method on DSC, IoU, Precision, and Recall are 72.7%, 62.3%, 85.8%, and 69.4%, respectively. Compared to the second-best method, the three indicators are improved by 1.9%, 2.4%, and 16.1%, respectively.

## 5. Discussion

The purpose of this paper is to present a segmentation algorithm using deep neural networks to segment medical images, especially BUSI. A summary of the results of the evaluation can be found in Table 2&5. According to the results of study, Proposed model was able to accurately detect boundaries in low-contrast images of breast ultrasounds and can segment breast tumors in breast ultrasound images. Thus, by accurately detecting tumor regions and boundaries, the next step is to correctly classify tumors as benign or malignant.

All results indicate that proposed model has improved the segmentation performance compared to other models considerably. As a result of our design, Co-block integrates both low-level and high-level features. Note that in order to improve the accuracy, Low-level features are needed, and in order to extract complex features, it is necessary to have high-level features. As a means of evaluating the performance of the proposed method on BUSI, DSC, IoU, ACC, and AUC were used. Proposed model improved the DSC, IoU, and ACC by 7.8%, 9.9%, and 2.4%, respectively.

Designed NN is tested on benign and malignant breast ultrasound images of BUSI separately to evaluate the robustness of the proposed method. DSC, IoU, Precision, and Recall were used to evaluate the segmentation performance of the proposed method on benign and malignant BUSI. As shown in Table 5, it can be seen that the proposed method has better segmentation performance on both benign and malignant BUSI compared to other deep learning models. A comparison of proposed network parameters with other models also shows that proposed network uses fewer parameters when compared with other models. As a result, we have a network that is lighter and requires less computation.

There are some qualitative results that can be seen in Fig. 7, which contains the segmentation results of both the proposed model and a variety of other models. It can be seen from Fig. 7 that the segmented images of proposed model match the ground-truth images well enough; the segmented images obtained from proposed model are very similar to the ground-truth images in terms of their segmentation. It can be seen from Fig. 7 that in some BUSI there are two tumors that are close to each other. With the help of proposed model, we are able to separate and segment them accurately.

## 6. Conclusion

This paper proposes an encoder-decoder U-Net based neural network to improve breast ultrasound image segmentation accuracy. Data augmentation was achieved through denoising and rotating the data during preprocessing. The NLM method is used to reduce speckle noise in breast ultrasound images. To increase the accuracy of the segmentation process, we designed a new block, Co-Block, which fuses both low-level and high-level extracted features, and integrated Res-Net's identity block and MultiResUnet's block into U-Net architecture to accomplish a better segmentation process. A set of experiments was conducted on the BUSI dataset in order to validate the designed network. Using the BUSI dataset, the segmentation performance of our network method was evaluated by comparing it with a variety of state-of-the-art deep learning segmentation methods. The experimental results demonstrated that proposed network outperforms U-Net, D-UNet, U-Net++, Seg-Net, RRCNet, and other deep networks in terms of segmentation accuracy. In the future, we will try to integrate the proposed network with the transformer network to obtain better segmentation accuracy for medical ultrasound images. It is also possible to classify segmented images into benign and malignant tumors using another neural network, and as a result, segmentation of images using the designed network can be used as a pre-processing for the next step, which is tumor classification.